\documentclass[prd,twocolumn,epsfig]{revtex4}
\newcommand{\be}{\begin{equation}}
\newcommand{\ee}{\end{equation}}
\newcommand{\<}{\langle}
\renewcommand{\>}{\rangle}
\newcommand{\reff}[1]{(\ref{#1})}
\newcommand{ \K }{ \raisebox{-0.8ex} {\scriptsize \it K} }

\begin{document}

\title{Composite boson dominance  in many-fermion systems}

%\author{Fabrizio Palumbo~\thanks{This work has been partially 
%  supported by EEC under the contract HPRN-CT-2000-00131}}
%\address{INFN -- Laboratori Nazionali di Frascati - P.~O.~Box 13, I-00044 Frascati, ITALIA}

\author{Fabrizio Palumbo}
 \altaffiliation[]{This work has been partially 
  supported by EEC under the contract MRTN-CT-2004-005104}%Lines break automatically or can be forced with} \\
\affiliation{%
  {\small\it INFN -- Laboratori Nazionali di Frascati}  \\%[-0.2cm]
  {\small\it P.~O.~Box 13, I-00044 Frascati, ITALIA}          \\%[-0.2cm]
  {\small e-mail: {\tt fabrizio.palumbo@lnf.infn.it}}     
}%

%\author{
%  { Fabrizio Palumbo~\thanks{This work has been partially 
%  supported by EEC under the contract HPRN-CT-2000-00131}}             \\%[-0.2cm]
%  {\small\it INFN -- Laboratori Nazionali di Frascati}  \\%[-0.2cm]
%  {\small\it P.~O.~Box 13, I-00044 Frascati, ITALIA}          \\%[-0.2cm]
%  {\small e-mail: {\tt fabrizio.palumbo@lnf.infn.it}}     
%   }

%\date{\today}

\thispagestyle{empty}   % Suppress page number on front page.

\begin{abstract}
I recently proposed  a method of bosonization based on the use of coherent states of fermion composites, whose validity was restricted
to smooth structure functions. In the present paper I remove this limitation and derive  results which hold for arbitrary
interactions and structure functions. The method respects all symmetries and in particular fermion number conservation. 
It reproduces exactly the results of the pairing model of atomic nuclei and of the BCS model of superconductivity 
in the number conserving form of the quasi-chemical equilibrium theory.

%\vspace{0.3cm}

%\noindent Keywords:

\end{abstract}

\maketitle

\clearpage

%\nopagebreak

\section{Introduction}

There are many finite and infinite fermion systems whose partition function at low energy is dominated by bosonic modes.
This is always the case when, due to spontaneous breaking of a global symmetry, there are Goldstone bosons.
The  effective bosons of these systems can be charged (fermion number 2), like Cooper pairs in metal, nuclear and so called 
color superconductivity\cite{Barr} or neutral (fermion number 0), like phonons in condensed matter and pions 
in hadronic physics.

The structure  of composite bosons can also be changed by  varying temperature or control parameters: for instance strongly 
interacting  fermionic atoms in magnetic traps can form at some temperature molecular pairs which  however condense only at a lower 
temperature. Moreover at zero temperature molecular pairs can be transformed into Cooper pairs by tuning an external magnetic field which
controls a Feshbach resonance. For such systems two features  are most relevant: the role of molecular versus Cooper 
pairs and the contribution of noncondensed pairs to the ground state energy~\cite{Staj}.

There is an immense literature  concerning techniques to derive effective Hamiltonians for
 both types of effective bosons. I will refer to all of them as bosonization methods. To put my work into perspective
 I will briefly review the most relevant step
 in the development of this subject, but with no attempt to completeness.

The first approach to bosonization of which I am aware  is due to Bogoliubov~\cite{Bogo}. 
After the BCS work on superconductivity Bogoliubov reformulated their theory  using 
the Fr\"ohlich Hamiltonian \cite{Froh} of electrons interacting with lattice phonons and rederived all their results concerning ground state properties.
He then mapped the Cooper pairs into effective bosons and found that their dynamics  is described by
the Hamiltonian of a superfluid system of elementary bosons he studied previously, whose  excitations are called Anderson-Bogoliubov sound.
These modes are related to  a nonvanishing contribution of noncondensed pairs to the ground state energy density. 

The presence of bosons external to the electron system, the lattice phonons,  plays an essential role in  Bogoliubov's theory. But in atomic 
nuclei in which nucleons are assumed to interact via a nucleon-nucleon potential (without mesons),  in gaseous systems of fermionic atoms 
and in the BCS model there are no external bosons. In particular in the BCS model the contribution of noncondensed pairs
to the ground state energy density  must vanish since the BCS 
solution for the ground state energy-density is exact in the thermodynamic limit~\cite{Haag}, and therefore collective excitations 
 cannot coincide with  the Anderson-Bogoliubov sound.

 The importance of superconductivity in atomic nuclei was immediately understood by Bohr Mottelson and Pines~\cite{Bohr} and
 the method of BCS, which breaks fermion number conservation  was adapted to atomic nuclei~\cite{Baym} for which this symmetry
  is important.
 Immediately superconductivity of infinite nuclear matter was investigated by many authors~\cite{Coop}, and 
  much later it was suggested that because of the strong tensor force Cooper pairs in symmetric  nuclear matter should 
  have the deuteron quantum numbers~\cite{Palu1}. Subsequently  a smooth transition from Bose-Einstein condensation of
  deuteron-like bound states at low densities and temperatures to BCS pairing at higher densities was considered~\cite{Stei}.
  
  More recently the superconducting properties of ultrasmall metallic grains have been investigated~\cite{Duke}. In this context, like
for atomic nuclei, it is important a theoretical method which respects fermion number conservation~\cite{Brau}.

   There have been many attempts 
  to reformulate the nuclear Hamiltonian in terms of
effective bosons. Beliaev and Zelewinski ~\cite{Beli} made an organic theory of bosonization with some points common to the present method,
but their expansion has problems of convergence and violates fermion number conservation. Later on much in the spirit of BCS but in a phenomenological 
approach Arima and Iachello~\cite{Arim}
introduced two different composite bosons, the $s-$ (spin 0) and $d-$ (spin 2) bosons. Their model, the Interacting Boson Model, proved
extremely successful in reproducing low energy nuclear properties, but it has not been derived in a
fully satisfactory way from a fundamental nuclear Hamiltonian.
 One of the features of the model to be understood  is why can one (mostly) restrict the boson space
to the $s$- and $d$-bosons. A justification will
emerge in the method I will present which, even though is found with a very specific fermion-fermion interaction, might
 have a more general validity. I must notice that the absence in the Interacting Boson Model of bosons with the deuteron 
quantum numbers predicted in  infinite nuclear matter might be due to the large size~\cite{Palu1} of the deuteron-like Cooper pairs
(very roughly) estimated of about 17 fm, and the fact that  in heavy atomic 
nuclei the valence shells of the protons are different from those of  neutrons.

In general there are many methods of bosonization. In 1+1 dimensions there is a wealth of exact results~\cite{Gogo}. 
These results have been extended~\cite{Hald} to many dimensions: first by use of renormalization group 
transformations
the fermion space is reduced to a tiny shell around the Fermi surface,  then the problem is essentially reduced  to one
dimension  by considering only excitations normal to the Fermi surface. In this way many general properties
can be investigated, but in many cases the effective parameters introduced by renormalization group transformations
are difficult to evaluate.

Different approach in multidimensional problems are based on several recipes ~\cite{Klei} for mapping of a fermion model
space into a boson space. Such methods respect fermion number conservation and in principle yield an exact
solution to the problem, but in practice one has to perform a truncation in the fermion space related to a selection of
 degrees of freedom guided by physical insight and calculational convenience. One shortcoming of this procedure
is the appearance of "intruders", namely states which in spite of their low energy do not appear in the boson space
generated by the mapping~\cite{Diep}.

For many-body systems like fermionic gases in magnetic traps the most common approaches are based on numerical simulations or
the quasi-chemical equilibrium theory~\cite{Scaf},\cite{Legg}, but a true bosonization to my knowledge has not been achieved.

I developed an approach to bosonization~\cite{Palu} in  which an  effective  bosonic action  is derived by evaluating 
the fermion partition function in a basis of coherent states of fermionic composites. Coherent states  offer for composites the same 
advantages they give for elementary bosons and fermions. 

 After bosonization is achieved the fermion dynamics can be studied by functional or numerical methods \cite{Kalo}.
  Analytic calculations are also possible in this approach both in the path integral and Hamiltonian formalisms,
at the price of  an expansion in the inverse of the dimension of the fermion space. Such an expansion respects fermion number conservation and therefore 
 can be used also for finite systems, and in fact its first application was to atomic nuclei.

An important issue was left over in the original paper. It concerns a subtraction necessary to perform the expansion in the presence of a 
condensate. The expansion was done under the assumption that the structure function of the condensed bosons be almost constant, a restriction
which can be fulfilled in some atomic nuclei but is not in most systems, including the BCS model. For this reason it was not 
possible to test the method on this theory. This limitation is removed in the present paper, and {\it general results are derived for arbitrary
fermion-fermion interactions and  structure functions of the composites}.  Actually the formalism is more general, as it will be 
illustrated
 at the end of Section III, where the effective bosonic action is reported. But let me anticipate
that it has been applied also to relativistic field theories, including gauge theories, at zero~\cite{Cara}
and finite fermion number~\cite{Paluf}. In the nonrelativistic domain the present method of bosonization finds  its
potential applications in finite systems like atomic nuclei and small metallic grains and in systems with nonseparable interactions
like BEC of fermion gases.

The paper is organized in the following way. To make it reasonably selfcontained I included some 
material from Refs.~\cite{Palu}, \cite{Cara}. In Sections 2 and 3 I outline 
 the approach and  the 
derivation of the effective bosonic action~\cite{Palu}, whose details  can be found in Appendix \ref{action}.
 Section 4 contains the derivation,  valid for
arbitrary fermion-fermion interaction, of the effective boson Hamiltonian which in Section 5 is expanded 
in the inverse of the dimension of the fermion space.  The leading order, given by Eq. \ref{H'},
 has the form of the Bogoliubov model of a superfluid boson system.  

As a test of the method for finite systems I apply it  in Section 6  to the pairing model of nucleons in
a single $j$-level.  {\it The  spectrum~\cite{Ring} is  correctly reproduced in a form
useful for an understanding and justification of the Interacting Boson Model of nuclear physics}.

Section 7 contains an application to infinite systems.  All the results of the BCS model of superconductivity concerning the ground state are  
exactly reproduced in  the number conserving form of the quasi-chemical equilibrium theory~\cite{Scaf}. 
 
In Section 8 I summarize the results and conclude with an outlook. 
 
 As already said  in the present paper I do not deal with neutral composites of particle-hole type, but I will introduce 
 some neutral  composites constructed with charged composites. Particle-hole  bosons will be discussed by an 
 extension of the method  in a future work. 
 
 At last I want to mention that the present approach has been applied also to relativistic field theories~\cite{Cara}, including
 gauge theories. 
 There are technical differences, due to ultraviolet divergencies in relativistic theories, which make necessary a different
 way of evaluating  the effective action of the composite bosons.
 But the essential strategy remains the same based on the use of coherent states of composites in the framework of the transfer matrix
 formalism which is close to the Hamiltonian formalism of nonrelativistic theories.  
 The method has been tested on  models of fermions with large number of flavors and  quartic  interactions  in 3+1 dimensions,
  exactly reproducing the gap equation 
 for spontaneous  breaking of a discrete chiral symmetry and the mass of the effective boson
 appearing in these models. Moreover the structure function of the condensed bosons has been determined for the first time. Its
 spatial part turns out, surprisingly enough, to be identical to that of the Cooper pairs of the BCS model.

\section{Outline of the approach}

Consider the partition function of a  system of elementary (non composite) bosons
\be
Z = \mbox{tr} \left[ \exp \left( - { 1 \over T} ( H_{\mbox{elem}} - \mu \, \hat{ n}) \right) \right]
\ee
where
$T$ is the temperature,  $\mu$ the chemical potential and $\hat{ n} $ the  number operator.
 A sector of $n$ particles can be selected by the constraint
\be
T {\partial \over \partial \mu} \ln Z = n.             \label{fermionnumber}
\ee
 A functional form of $Z$ can be found by performing the trace over coherent states~\cite{Nege}
\begin{equation}
|\alpha \rangle = \exp \left( \sum_K  \hat{\alpha}_K^{\dagger}{\hat \alpha}_K\right) |0 \rangle \,,
\end{equation}
where $K$ are the particles quantum numbers, $ \hat{\alpha}_K^{\dagger} $ their canonical creation operators and $\alpha_K $ holomorphic
 variables. Coherent states satisfy the basic or defining equations
\begin{equation}
\hat{\alpha}_K |\alpha \rangle = \alpha_K |\alpha \rangle
\end{equation}
where $\hat{\alpha}_K  $ are canonical destruction operators. In terms of these states we can write the identity in the Fock space 
\begin{equation}
\mathcal{I} = \int d\mu(\alpha^*, \alpha) \langle \alpha|\alpha \rangle^{-1} | \alpha \rangle \langle \alpha |
\end{equation}
where 
\be
d\mu(\alpha^*, \alpha) =   \prod_K \left[{d\alpha_K^* d\alpha_K \over 2\pi i}\right]  \,.
\ee
Using this resolution of the identity the trace in the partition function can be evaluated with the result~\cite{Nege}
\begin{equation}
Z = \int d\mu(\alpha^*, \alpha) \exp \left[- S_{\mbox{elem}}(\alpha^*, \alpha ) \right]\,.
\end{equation}
In the above equation $S_{\mbox{elem}} $ is the action of the particles
\begin{eqnarray}
S_{\mbox{elem}}& = &\tau \sum_t \left\{ - \alpha_t ^* \nabla_t \alpha_{t-1}  + H_{\mbox{elem}} (\alpha_t^*,\alpha_{t-1}) \right.
\nonumber\\
 & & \left. - \mu  \, \alpha_t^* \alpha_{t-1} \right\} \,, \label{actelem}
\end{eqnarray}
where 
\be
\nabla_t \, f = { 1 \over \tau} \left( f_{t+1} - f_t \right)  ,
\ee
and  $ H_{\mbox{elem}}(\hat{a}^{\dagger}, \, \hat{a})$ is the Hamiltonian  in normal order.

In a system of fermions whose low energy excitations are dominated by fermion composites  I can restrict the trace to 
these composites. The restricted partition function can be written
 \be
Z_C = \mbox{tr} \left[ {\mathcal P}_C \exp \left( - { 1 \over T} (H_{F}- \mu_{F}  \, \hat{ n}_{F})  \right) \right]
\ee
where ${\mathcal P}_C$ is a projection operator in the subspace of the composites. By analogy to elementary bosons 
I assume for it the approximate expression
\be 
{\mathcal P} = \int {d\beta^* \,d  \beta \over 2 \pi i  } \, \< \beta| \beta\>^{-1} |\beta \>\<\beta|  \label{Poperator}
\ee
where $  \beta^*,  \beta  $ are holomorfic variables and $|\beta \>$  coherent states of composites
\be
|\beta \> = |\exp \left( \sum_J \beta_J {\hat B}^{\dagger}_J \right) \>.
\ee
 The  creation operators  of composites with fermion number 2 are
\be
{\hat B}^{\dagger}_J 
= { 1\over 2\sqrt{\Omega_J}} \sum_{m_1,m_2} {\hat c}^{\dagger}_{m_1}\left( B^{\dagger}_J\right)_{m_1,m_2}
{\hat  c}^{\dagger}_{m_2}\,,
\ee
while those of fermion number zero are
\be
\hat{\Phi}^{\dagger}_J =  { 1 \over  \sqrt{\Omega_J}}\sum_{m_1 m_2}   {\hat c}_{m_1}^{\dagger}
 \left(\Phi^{\dagger}_J\right)_ {m_1 m_2}{\hat c}_{m_2}\,.
\ee
The ${\hat c}^{\dagger}$'s are fermion creation operators, $m$ and $J$, represent all the fermion and boson quantum numbers
and $\Omega_J$ is the index of nilpotency of the $J$-composite, which is defined as the largest integer such that
\be
\left( \hat{B}_J \right)^{\Omega_J} \neq 0. 
\ee
 I will assume for simplicity the index of nilpotency equal to half the dimension of the fermion space
for all the  composites.
 The matrices $B_J, \Phi_J$ are {\it the structure functions of the composites, which must be determined by a variational 
 calculation}. 
  Only the solutions with high index of nilpotency are acceptable. 
  
  I call the states $|\beta \rangle$  coherent  because they share with coherent states of elementary bosons
the property of a fixed phase relation among the components with different number of composites. 
But the basic property of coherent states cannot be fulfilled. Indeed
\begin{equation}
\hat{B}_J | \beta \rangle \neq \, \beta_J | \beta\rangle\,.
\end{equation}
This is a consequence of the composites commutation relations, which are not canonical
\begin{equation}
\left[ \hat{B}_J ,   \hat{B}^{\dagger}_K  \right] = { 1 \over 2}
 \mbox{Tr} \, (B_J  B^{\dagger}_K) 
- {\hat c}^{\dagger}B^{\dagger}_K B_J \,{\hat  c }\,. \label{comm}
\end{equation}
In  states with a number of composites  $ n << \Omega$, the above equations can be  approximately satisfied
provided the structure functions are sufficiently smooth.
Indeed in such a case the last  term is of order $n / \Omega $. But in  states with  $n  \sim \Omega$, it is not possible to satisfy 
them even with an absolute freedom about the form of the structure functions (which are instead determined by
the dynamics). The best we can do~\cite{Palu} is to satisfy them for states with $n+k$ composites, for fixed $n\sim \Omega$ and
 $|k| << \Omega$. 

 As stated in the Introduction, in the present work I will study only composites with fermion number 2.  The properties 
 of the operator ${\mathcal P} $  are reported in  Appendix \ref{projector} in the form derived in Ref.~\cite{Cara}. 
  
The trace can be exactly evaluated~\cite{Palu} (see below) yielding a functional form of $Z_C$
\be
Z_C=\int \left[ { d \beta^* d \beta \over 2 \pi i} \right]\exp \left( - S_{\mbox{eff}}(\beta^*, \beta) \right).
\ee
The expression of the effective bosonic action $S_{\mbox{eff}} $ is reported in the next Section. 
This result holds under the only physical assumption of boson dominance and the approximation adopted for ${\mathcal P} $. 

In many-body physics it is often used  the Hamiltonian formalism. The Hamiltonian of the effective bosons, $H_B$,
cannot be read directly from the effective action, because $S_{\mbox{eff}} (\beta^*,\beta)  $ does not have the form
of an action of elementary bosons. Indeed it contains anomalous time derivative terms, anomalous couplings of the
chemical potential and nonpolynomial interactions, which are all features of compositeness. Therefore it has
been necessary to devise an appropriate procedure to derive $H_B$, which is given in terms of boson operators
 ${\hat b}^{\dagger}, {\hat b}$, 
(not to be confused with the composite operators  ${\hat B}^{\dagger}, {\hat B}$) satisfying canonical commutation relations, so that
\be
Z_C = \mbox{tr} \left( -{ 1 \over T} (H_B - \mu_B \hat{ n}) \right).
\ee
 $\mu_B$ is the boson chemical potential and $\hat{ n} $ the boson number operator.

$H_{B}$ has a closed form but, for a practical use, it is necessary to perform an expansion 
in  inverse powers of the index of nilpotency $\Omega$.

\section{The effective bosonic action}

 The most general fermion-fermion interaction  can be written as a sum of separable terms, so that the fermion 
 Hamiltonian can be given the form
\begin{eqnarray}
H_{F} &=&{\hat  c}^{\dagger} h_0 \, {\hat c} -
\sum_K g\K  \, { 1\over 2} \, {\hat c}^{\dagger} F_K^{\dagger}{\hat c}^{\dagger} \, 
{ 1\over 2}\, {\hat c} \, F_K \,{\hat  c}. \label{Hami}
\end{eqnarray}
 The one-body term includes the single-particle energy 
with matrix $e$, the fermion chemical potential $\mu_{F}$ and any single-particle interaction with external fields
included in the matrix ${\cal M}$  
\be
h_0= e - \mu_{F}+ {\cal M} .
\ee
The matrices $F_K $ are the form factors of the potential, normalized according to
\be
 \mbox{tr} ( F_{K_1}^{\dagger} F_{K_2}) = 2 \, \Omega \delta_{K_1 K_2}.  \label{potnormal}
\ee
In order to evaluate $Z_C$ I divide the inverse temperature in $N_0$ intervals of size $\tau$
\be 
T = { 1 \over N_0 \tau}.
\ee
Then as shown in Appendix \ref{action} the  Euclidean effective action is
 \begin{eqnarray}
& & S_{\mbox{eff}}(\beta^*,\beta) =\tau \sum_t { 1 \over 2} \mbox{tr} \left\{ { 1\over  \tau}   \ln \left[
1\!\!1 + 
 \tau \, R \, {\mathcal B}^{\dagger} \nabla_t {\mathcal B} \right] \right.
\nonumber\\
& & \left.
+   2  R  \,{\mathcal B}^{\dagger}   h \, {\mathcal B}    - 
 \sum_K g_K  \left[ (R -1) \,  F^{\dagger}_K F_K  \right. \right.
\nonumber\\
& & \left. \left.
 +  (R {\mathcal B}^{\dagger}  F_K^{\dagger}) \,
{ 1 \over 2} \mbox{tr}(R F_K {\mathcal B}) 
-   R {\mathcal B}^{\dagger} F_K^{\dagger} R  F_K {\mathcal B}  \right]
 \right\} 
\label{bosaction}  
\end{eqnarray}
where
\be
h = h_0 -  \sum_K g\K   F_K^{\dagger} F_K  \label{hoperator}
\ee
\be
{\mathcal B} = {1 \over \sqrt{\Omega}} \sum_J \beta_J B_J^{\dagger} = {1 \over \sqrt{\Omega}}  \beta \cdot
B^{\dagger} 
\ee
\be
R= \left( 1\!\!1 + {\mathcal B}^{\dagger} {\mathcal B}\right)^{-1}  \,.
\ee
 Notice in the third line a trace inside the trace. {\it The variables $\beta^*, 
\beta$  are always understood at times $t,t-1$ respectively}.  Explicitly, for instance
\be
R_t= \left( 1\!\!1 + {\mathcal B}_t^{\dagger} {\mathcal B}_{t-1}\right)^{-1}.
\nonumber\\  
\ee
$S_{\mbox{eff}} $ has a global $U(1)$ symmetry which implies boson conservation.

 The fermionic interactions with external fields appear in the
bosonic terms which involve the matrix ${\cal M}$ (appearing in $h$). 

The dynamical problem of the interacting (composite) bosons can be
solved within the path integral formalism. Part of this problem is the determination of the structure matrices $B_J$ which 
 can be done by a variational calculation. The present approach then shares two important features with
 variational methods: the restriction of the fermion space to a subspace, the space of the composites, and the
 variational determination of the structure functions. But unlike standard variational methods excited states 
 are treated at the same time and on the same footing as the ground state.
 
 Before showing a concrete way of application, I want to list some characteristic features of the present formalism:

1) It can be used with interactions quartic as well quadratic in the fermionic fields. The latter ones include electromagnetic 
and phonon interactions in the nonrelativistic domain, and renormalizable relativistic field theories in 3+1
dimensions~\cite{Cara, Paluf}. In the application to these theories the main change is the replacement 
of the exponential of the Hamiltonian by the transfer matrix

2) It can be extended to cases in which not all the fermions bosonize, like  odd atomic nuclei, systems
of fermions and composite bosons in quasichemical equilibrium, and relativistic field theories
 at finite fermion density. This latter case has already been studied for QCD at finite temperature and baryon density~\cite{Paluf}, 
and one can see that the procedure adopted can easily be applied to nonrelativistic systems as well

3) It allows the treatment of  different coexisting composite bosons, for
instance Cooper and molecular pairs. In such a case the equations for the structure functions of the composites 
should have more solutions, possibly with the same quantum numbers

4) It provides the structure functions of the composites, and therefore their effective coupling to other fields,
like the electromagnetic field

5) It allows the introduction of composites with quantum numbers different from those of the form factors of the potential.
An example is given in Section VI: with a purely pairing interaction in atomic nuclei, one needs
also bosons with angular momentum different from zero.

\section{The effective boson Hamiltonian}

As already said the derivation of the boson Hamiltonian is not straightforward, because $S_{\mbox{eff}}$ differs in many respects from
the action of elementary bosons shown in Eq.(\ref{actelem}). 
I notice  that in $ S_{\mbox{eff}}$ 

i) the time derivative term is not canonical

ii) the coupling of the chemical potential (appearing in $h$) is also noncanonical 

iii) there are non polynomial interactions because of the $R$-function. This function  becomes singular,
 as it will become clear in the sequel, when the number of bosons is of order $\Omega$, reflecting the Pauli
principle.

Let us start by examining  the features of compositeness when the number of bosons is much smaller than $\Omega$. Since
the expectation  value of $\beta^* \cdot \beta $ is of the order of the number of bosons, in this case we can perform an
expansion of logarithm and  $R$-function in inverse powers of $\Omega$. From the logarithm I get
\begin{eqnarray}
& & { 1\over 2 \tau} \,  \mbox{tr} \ln \left[ 1\!\!1 + 
 \tau R {\mathcal B}^{\dagger} \nabla_t {\mathcal B} \right] = 
{ 1\over 2} \mbox{tr} \left({\mathcal B}^{\dagger} \nabla_t {\mathcal B} \right)
\nonumber\\
& &  - { 1\over 4 } \mbox{tr}  \left[ {\mathcal B}^{\dagger} {\mathcal B} \, {\mathcal B}^{\dagger} \nabla_t{\mathcal B} \right] +... 
\end{eqnarray}
The first term can be made canonical by normalizing the boson form factors according to
 \be
\mbox{tr} (B^{\dagger}_J \,  B_K ) = 2 \Omega \, \delta_{J,K}. \label{norm}
\ee
The other terms are then of order $\Omega^{-1}$.
Notice that the diagonal condition is only a matter of normalization, but the off diagonal one must be compatible
with the dynamics, and a redefinition of the $\beta$'s can be  necessary to get it. 

Expanding the $R$-function I get the following couplings of the fermion chemical potential
\be
\mu_{F}  \, \mbox{tr} \left[ {\mathcal B}_t^{\dagger} {\mathcal B}_{t-1} - { 1 \over 2} ( {\mathcal B}_t^{\dagger}  {\mathcal B}_{t-1})^2 +...\right]. \label{noncan}
\ee
Only the first term is canonical. However
 the anomalous couplings can be eliminated to order $\Omega^{-1}$ by a redefinition of the chemical
potential~\cite{Palu}, so that in the case of a small number of bosons the Hamiltonian can be derived without difficulty.

But when the number of bosons is of order $\Omega$, an expansion of logarithm and $R$-function
 can be performed only after an appropriate subtraction, as explained in Section V. I assume, and I will verify later, that after such
a subtraction the anomalous time derivative terms  be of order $ \Omega^{-1}$.
Then $Z_C$ can be written in terms of an auxiliary Hamiltonian $H'$ as a trace in a boson space 
\be
Z_C= \mbox{tr} \exp \left( - {1 \over T} H'\right).
\ee
 $H'$ is obtained  from $S_{\mbox{eff}}$ by omitting the time derivative term,
and replacing the variables $\beta^*,\beta $ by corresponding creation-annihilation operators
${\hat b}^{\dagger},{\hat  b}$. These  satisfy canonical commutation relations and should not be confused with the
corresponding operators ${\hat B}^{\dagger},{\hat  B}$ for the composites
\begin{eqnarray}
& & H' = \,\, :{ 1 \over 2} \mbox{tr}\left\{ 
2{\hat R}  \, {\hat {\mathcal B}}^{\dagger}   h \,\, {\hat {\mathcal B}}  
 -  \sum_K g_K    \left[   \left({\hat R} -1 \right) \,   F^{\dagger}_K F_K 
   \right. \right.
 \nonumber\\
 & &
    \left.   \left.\, + 
     {\hat R} \,{\hat {\mathcal B}}^{\dagger}  F_K^{\dagger} 
 { 1 \over 2} \mbox{tr}({\hat R} \, F_K {\hat {\mathcal B}}) 
 -   {\hat R} \, {\hat {\mathcal B}}^{\dagger} F_K^{\dagger} {\hat R} \, F_K {\hat {\mathcal B}}   
  \right] 
\right\}:. 
\nonumber\\   
\end{eqnarray}
The colons denote normal ordering and  
\begin{eqnarray}
{\hat {\mathcal B}} & = & {1\over \sqrt{\Omega}}\,  {\hat b} \cdot B^{\dagger}
\nonumber\\
{\hat R} &=&
\left[ 1\!\!1 + { 1 \over \Omega} \,
{\hat b}^{\dagger} \cdot B \,\,{\hat b }\cdot B^{\dagger}  \right]^{-1}.  \label{gammab}
\end{eqnarray}
From $H'$ I will derive in the next Section  the boson Hamiltonian $H_{B}$, Eq. (\ref{bosham}).
I must notice that the case of a number of bosons much smaller than $\Omega$ cannot be retrieved from the above equations. The reason is that 
in the former case $\mu_{F} = O(\Omega) $, while $\mu_{F}= O(\Omega^{0})$ in the case $n \sim \Omega $.

  \section{The $\Omega^{-1}$ expansion}
  
  A rather general way to perform the subtraction necessary for the $\Omega^{-1}$ expansion  is to write the operator 
  ${\hat R}$ in the form
  \be
{\hat R} = ( 1\!\!1 + {\hat \eta})^{-1}  \, \Gamma
\ee
where
\begin{eqnarray}
{\hat \eta} &=&  { 1\over \Omega} \, \Gamma    
 \sum_{K_{1 }K_{2}} \left( {\hat  b}^{\dagger}_{K_1}{\hat  b}_{K_2} - r_{K_{1 }K_{2}}^{2} \right)  B_{K_{1}}B^{\dagger}_{K_{2}}
 \nonumber\\
 \Gamma &=& \left[  1\!\!1 +  { 1\over \Omega} \,  
 \sum_{K_{1 }K_{2}}  r_{K_{1 }K_{2}}^{2}  B_{K_{1}}B^{\dagger}_{K_{2}}  \right]^{-1} .
\end{eqnarray}
Notice that  ${\hat R} $ can be expanded with respect to ${\hat \eta}$, while $\Gamma$, which  however does not contain creation-annihilation operators,
 must be treated exactly. 
The parameters $r^{2}_{K_{1} K_{2}}$ are related to the expectation values $\< {\hat b}_{K_{1}}^{\dagger} {\hat b}_{K_{2}}\> $.
It is important to observe that such  expectation values  do not break boson  (and therefore fermion) number
conservation, but will in general break other symmetries, like rotational invariance in deformed atomic nuclei. Their determination
allows therefore to study thermodynamic or quantum phase transitions like breaking of rotational symmetry by axial or triaxial shapes in atomic nuclei~\cite{Iac2}.

For the sake of simplicity  in ~\cite{Palu} I restricted myself to cases in which  the  structure functions are almost constant, which justifies
 a subtraction independent of them. The formalism was then tested in the case of the so called pairing model~\cite{Ring}. The ground state energy  
 was exactly reproduced, 
 but the spectrum of excitations was not studied because the boson Hamiltonian  contains couplings of all the bosons among themselves and was not 
 solved. 
 As we will see such couplings are an artifact due to the inadequacy of that subtraction.

\subsection{Subtraction in the presence of an $s$-condensate}

In the present paper I remove the  restriction that the form factors should be almost constant, and consider the case in which a condensation occurs in a single quantum mode,
called the $s$ mode, whose quantum numbers will be denoted by "zero"
\be
r^{2}_{K_{1} K_{2}}= \delta_{K_{1},0} \, \delta_{K_{2},0} \, r^2 . \label{subtraction}
\ee
 By condensation of the $s$-boson I understand  that the occupation number of this 
mode is of order $\Omega$. The  terms appearing in ${\hat \eta} $  can then be classified according to
\begin{eqnarray}
& &    ({\hat b}_{0}^{\dagger}{\hat b}_{0} - r^{2}) B_{0}^{\dagger} B_{0} + \sum_{K_1,K_2 \ne 0}
{\hat b}_{K_{1}}^{\dagger}  {\hat b}_{{K}_2} \,  B_{{K}_1} \, B^{\dagger}_{{K}_2} 
\,\, \sim \,\,\, \Omega^0
\nonumber\\
& & {\hat b}_0^{\dagger} \,  B_0 \sum_{K \ne 0}
 {\hat b}_K \, B^{\dagger}_K \,\,\, \sim \,\,\,\sqrt{\Omega},  \label{class}
\end{eqnarray}
provided  the sum of the occupation numbers of noncondensed  modes is much smaller than $\Omega$ 
and the structure matrices are of order $\Omega^0 $.  Under these conditions one subtraction is sufficient, otherwise more subtractions are needed. 
Expansion of time derivative terms in $S_{\mbox {eff}}$ according to this classification, neglecting contributions of  order $ \Omega^{-1}$ gives
\begin{eqnarray}
& & { 1 \over 2 \tau} \mbox{tr}  \left[ \ln \left(1 + {\mathcal B}_{t}^{\dagger} {\mathcal B}_{t } \right) - 
 \ln \left(1 + {\mathcal B}_{t}^{\dagger} {\mathcal B}_{t-1 } \right)\right] =
 \nonumber\\
 & & \,\,\,\,\, \,\,\,\, { 1 \over 2 \Omega} \mbox{tr}\,  C_{00} \, \beta_{0}^{*}\nabla_{t}\beta_{0}
 + \sum_{K_{1}K_{2}}{ 1 \over 2 \Omega} \mbox{tr} \left[C_{K_{1}K_{2}} \right.
\nonumber\\
& &  \,\,\,\,\,\,\,\,\,  \left.   - { 1 \over \Omega} \beta_{0}^{*} \beta_{0} \, C_{K_{1}0} C_{0K_{2}} \right]
\beta_{K_{1}}^{*}\nabla_{t}\beta_{K_{2}}
\end{eqnarray}
where
\be
C_{K_{1}K_{2}}= \Gamma \, B_{K_{1}}B^{\dagger}_{K_{2}}.
\ee
In the derivation of the above equation I assumed   $\mbox{tr} \, C_{K 0} =0, \, \mbox{for} \, K \ne 0 $ and I 
disregarded  the fluctuations of the product $ \beta_{0}^{*} \beta_{0}^{*} $, whose contribution is of order $\Omega^{-1}$, so that terms of the form
\be
{ 1 \over 2 \Omega^{2}}  \mbox{tr}  ( C_{0K_{1}}C_{0K_{2}}) 
\beta_{0}^{*} \beta_{0}^{*} \nabla_{t}(\beta_{K_{1}}\beta_{K_{2}})
\ee 
 are total time derivatives and do not contribute to the action.   Then replacing $\beta_{0}^{*}\beta_{0}$   by $n$, the number of  bosons, the temporal
 terms become canonical if I impose the normalizations                                                                                                                   
\begin{eqnarray}                                                                                                                                                                                
& &{ 1 \over 2 \Omega } \mbox{tr} \,C_{00}  = 1, 
\nonumber\\
& & { 1 \over 2 \Omega} \mbox{tr} \left[ C_{K_{1}K_{2}}
 -  {n \over \Omega} C_{K_{1}0}C_{0K_{2}}\right]  =\delta_{K_{1}K_{2}}, 
  \nonumber\\
  & & \,\,\,\,\,\,\,\,\,\,\,\,\,    K_{1},K_{2}  \ne 0.          \label{ffnormal}
\end{eqnarray} 
It is worth while noticing that the ground state wave function of free fermions is included in the class of states which are assumed to dominate the partition function,
so that the variational evaluation of the structure function will tell whether an actual condensation will occur or not. Indeed if I choose
\be
(B_{0} B_{0}^{\dagger})_{m_{1} m_{2}}= \xi \, \delta_{m_{1}m_{2}}\theta(m_{F}-m_{1})
\ee
 where $\theta$ is the step function and $m_{F}$ are the fermion quantum numbers at the Fermi surface, for large $\xi$ 
the normalization gives
\be
r^{2}= { 1 \over 2} n_{F},
\ee
and  all the coherent states  are dominated by the term with $n_{F}$ fermions
\be
| \beta \> \sim ( \beta  \, \xi)^{{1\over 2}n_{F}}\prod_{m=1}^{m_{F}}c_{m}^{\dagger}.
\ee

\subsection{$\Omega^{-1}$ expansion in the presence of an $s$ condensate}

To proceed with the expansion of the boson Hamiltonian  it is necessary to 
 know how the coupling constants $g_{K}$ scale
with $\Omega$.  For infinite systems,  since $ \Omega \rightarrow \infty $, to get finite energies we must require
$ g_K \sim \Omega^{-1}$.  Such a behavior is also acceptable for many finite systems and I assume it in the following.
 As a consequence the fermion chemical potential $\mu_{F} $ is of order $\Omega^{0}$.
 
 At this point it is convenient to introduce the notations
 \be
 \nu = {r^{2} \over \Omega},  \,\,\,\,\,
 \hat{n}_{0}=  {\hat b}_{0}^{\dagger}{\hat b}_{0}.
   \ee
 Neglecting contributions of  order $ \Omega^{-1}$ I get
 \begin{eqnarray} 
  H'  &  \sim & E_{C} \,  + : \left\{  {\cal E}_{0} \,\hat{n}_{0} + { 1 \over 2}\sum_{K_{1}, K_{2} \ne 0}  
   \left[ \left({\cal E}_{K_{1}K_{2}} \,{\hat  b}_{K_{1}}^{\dagger}{\hat b}_{K_{2}} \right. \right. \right.
  \nonumber\\
  &   & \left. \left. \left. 
+  { 1 \over \Omega}  {\cal V}_{K_{1}, K_{2}}  \, {\hat b}_{K_{1}}^{\dagger} {\hat b}_{K_{2}}^{\dagger} 
{\hat b}_{0}{\hat b}_{0} \right) + H.c. \right]
 \right\}:   , \label{H'}
  \end{eqnarray}
where  $E_{C}$ is a c-number  and $H.c.$ represents the Hermitian conjugate of the operator in the round brackets.
The coefficients ${\cal E}$ and $ {\cal V}$ are functions of the operator
 $\hat{n}_{0}$, which explains the presence of  normal ordering.
 
 In many cases the double sum over $K_{1},K_{2}$ reduces to a single sum, $K_{2}$
  resulting conjugate to $K_{1}$ according to a one-to-one correspondence $K_{2} = \tilde{K}_{1}$
  \begin{eqnarray} 
  H'  & & \sim  E_{C} \,  + : \left\{  {\cal E}_{0} \,\hat{n}_{0} + \sum_{K \ne 0}  
   \left[{\cal E}_{K}{\hat b}_{K}^{\dagger}{\hat b}_{K} \right. \right.
  \nonumber\\
  & & \left. \left.
+  { 1 \over  2  \, \Omega}  {\cal V}_{K} {\hat b}_{K}^{\dagger} {\hat b}_{\tilde{K}}^{\dagger}
{\hat b}_{0} {\hat b}_{0}  + H.c. \right]
 \right\}:   ,
  \end{eqnarray}
   From their general expression,  reported in  Appendix \ref{coefficients}, we see
 that  $E_{C} \sim \Omega$ while the operators ${\cal E}_{K}, {\cal V}_{K}$ take values of order $\Omega^{0}$. 
  
  $H'$ has the form of the Bogoliubov model of superfluidity, with an important qualification to be discussed below.
Because of the absence of terms involving three or four operators of bosons out of the condensate, 
  $H'$ can be approximately (see below) diagonalized  introducing the phonon operators
  \be
{\hat A}_{K} = { 1\over \sqrt{n}} {\hat b}_{K}{\hat b}_{0}^{\dagger},  \,\,\,
{\hat A}_{K}^{\dagger} = { 1\over  \sqrt{n}} {\hat b}_{K}^{\dagger}{\hat b}_{0}.
\ee
 In terms of these  operators 
\be
H' = E_{C} \, + \delta E_{C} \, + : {\cal E}_{0}\hat{n}_{0} :+ \sum_{K \ne 0} E_{K}{\hat A}_{K}^{\dagger}
{\hat A}_{K}
\ee
where
\begin{eqnarray}
\delta E_{C} &= & { 1 \over 2} \sum_{K\ne 0}\left( E_{K}-  {\cal E}_{K}\right)
\nonumber\\
E_{K} &= & \sqrt{ {\cal E}_{K}^{2} - \left(  { n \over \Omega} \, {\cal V}_{K} \right)^{2}}.
\end{eqnarray}
The average number of noncondensed bosons in the ground state is~\cite{Marc}
\be
\< \sum_{K \ne 0}{\hat b}_{K}^{\dagger}{\hat b}_{K}\> = { 1 \over 2}\sum_{K \ne 0}\left( { {\cal E}_{K} \over  E_{K}} - 1 \right).
\ee
Approximating the commutation relations of the phonon operators by canonical ones  introduces errors of order $\Omega^{-1}$ only if this expectation value is much
smaller than $\Omega$.  This condition is also necessary for the classification of Eq.(\ref{class}) to hold.  Now even though the operators ${\cal E}_{K}$
 and $ {\cal V}_{K}$ take values of order $\Omega^{0}$, so that also the phonon energies $E_{K}$ 
 are of this order,  it is well possible that 
  $\< \sum_{K \ne 0}{\hat b}_{K}^{\dagger}{\hat b}_{K}\> = O(\Omega)$. {\it In such a case a unique subtraction is not sufficient and farther subtractions are necessary}.

Now the qualification mentioned above. As reminded in the Introduction  in his reformulation of the theory of superconductivity starting from the 
Fr\"ohlich Hamiltonian~\cite{Froh} of electrons interacting with lattice phonons, Bogoliubov
 found an effective Hamiltonian essentially equal to that he studied previously for superfluid bosonic systems as far as the bosonic excitations are concerned. The
 coefficients ${\cal E}_{K}$ and $ {\cal V}_{K}$ of this Hamiltonian are such~\cite{Marc} that the excitation spectrum is
  \be
  E_{K}\sim  |K|,\,\,\,\, |K| \rightarrow 0.
  \ee
 As a consequence  $\< \sum_{K \ne 0}{\hat b}_{K}^{\dagger}{\hat b}_{K}\>, \delta E_{C} = O (\Omega)$. 
 Such a result cannot hold for the  BCS model, because  the expression of the ground state energy density  derived by BCS is 
 exact~\cite{Haag}.

  \subsection{Determination of subtraction parameter and fermion chemical potential}

In the determination of the subtraction parameters I meet with a subtlety. $H'$ commutes with the boson number operator, so I can select sectors with
a given number of bosons. But I am not guaranteed that these bosons carry fermion number 2, because of
the noncanonical coupling of the chemical potential. This fundamental property can be enforced just exploiting the parameters introduced by the subtractions.
Indeed,  denoting  by $E'_0(n)$ the lowest eigenvalue of $H'$ in the sector  of $n$ bosons, I require that $E'_0(n)$  be the lowest eigenvalue for
$n ={ 1 \over 2 }\, n_F$
\be
{\partial \over \partial n} E_0'(n) =0, \,\,\,\mbox{for}\, \,  n = { 1 \over 2}n_F. \label{fermionnumber1}
\ee
 The above equation, together with the condition \reff{fermionnumber} on the fermion number,  determines one of the parameters $r$ and the fermion chemical 
 potential $\mu_{F}$ 
 as functions of the number of bosons, $\overline{r}= \overline{r}(n), \overline{\mu}_F = \overline{\mu}_F(n)$, ensuring that these bosons carry 
 fermion number 2.
The boson Hamiltonian in the sector of n bosons is finally
\be
H_B(n)= H'(\overline{r}, \overline{\mu}_F) + 2 \, \overline{\mu}_F \,n.  \label{bosham}
\ee
It depends  on $n$ explicitly and through the dependence on $n$ of  $\overline{r},\overline{\mu}_F$. Therefore
also the matrices $B_J$, the form factors of the bosons, will depend on $n$, namely on the number of fermions.

Notice that $H'$ provides a mapping of the fermion interactions with external fields 
\be
c^{\dagger}{\cal M}c \rightarrow  :  { 1 \over 2} \mbox{tr} \, \left\{ {\hat \Gamma} 
 \left[ {\hat {\mathcal B}}^{\dagger}  {\cal M}  {\hat {\mathcal B}}  
+ { 1\over 2}[ {\hat {\mathcal B}}^{\dagger}{\hat {\mathcal B}},{\cal M}]_{+} \,\right] \right\}: \,\,  .
\ee
The ground state energy of the auxiliary Hamiltonian in a state of $n$  bosons  is
\be
E'_{0}(n) = E_{C} \, + \delta E_{C} \, + \< n ,o| : {\cal E}_{0}(\hat{n}_{0}) \, \hat{n}_{0} : |n, o\> .
\ee
Taking normal ordering into account
\be
 \< n ,o| : {\cal E}_{0}(\hat{n}_{0}) \, \hat{n}_{0} : |n, o\> = \overline{{\cal E}}_0(n_0) \, n_0
 \ee
 where
 \be
 \overline{{\cal E}}_0(n_0) = {\cal E}_0(n_0) + 2  n_0 g_0 D_{00} D_{0000}.
 \ee
 $n_0$ is the number of condensed bosons in a state of $n$ bosons and 
 the quantities $D_{K_{1}K_{2}...}$ are defined in  Appendix B.
 The constraints \reff{fermionnumber} and \reff{fermionnumber1} 
 for subtraction parameter and chemical potential are
\begin{eqnarray}
{\partial  E'_{0}(n) \over \partial \mu_{F}} &=& -2 \left\{ r^{2} { 1 \over 2 \Omega}\mbox{tr} \, C_{00} 
+ \left( n_{0} - r^{2} \right)
{ 1 \over 2 \Omega}\mbox{tr} \left[  \Gamma C_{00} \right] \right\}
\nonumber\\
&+&  {\partial  \delta E_{C} \over \partial \mu_{F}} +  2 n = 0
\nonumber\\
{\partial  E'_{0}(n) \over \partial n} &=& {\partial  \over \partial n} \left[  \overline{{\cal E}}_0 (n_0) \,  n_0 \right] + 
{\partial  \delta E_{C} \over \partial n} =0.  \label{fermionnumber2}
\end{eqnarray}
I have two equations and three unknowns: $\mu_F, r^2, n_0$. But I remind that the $\Omega^{-1}$ expansion holds under the  assumption
\be
n - n_0 =  O(\Omega^0),
\ee
otherwise further subtractions are necessary. I assume, and I will discuss later when this assumption holds true
\be
{\partial  \over \partial \mu_{F}}  \delta E_{C}= O( \Omega^0),
\,\,\,\,\, 
{\partial   \over \partial n} \delta  E_C = O(\Omega^{-1}).
\ee
Then  the first equation with the normalization condition~(\ref{ffnormal}) gives
\be
\overline{r}^2 = n + O(\Omega^0), \label{subtractionparameter}
\ee   
 and using this result in the second equation I get
\begin{eqnarray}
{ 1 \over 2 \Omega} \mbox{tr} \left[ \Gamma^{2}B_{0}( e - \mu_F) B_{0}^{\dagger} \right] +
 2 \nu\, \Omega g_{0} D_{00}D_{0000} & &
 \nonumber\\
  -  \Omega g_{0} D_{00}^{2} = O(\Omega^{-1}) &&  \label{fermionchemical}
\end{eqnarray}
which determines $\overline{\mu}_{F}$. 

I can finally write the ground state eigenvalue of $H_B$
\begin{eqnarray}
E_{0}(n) &= &E'_{0}(n) +2n \overline{\mu}_{F} = \left\{ { 1 \over  \Omega} \mbox{tr} \left[ \Gamma B_{0}\, e
B_{0}^{\dagger}  \right] \right.
\nonumber\\
 & &  \left. \phantom{{1\over \Omega}}-\Omega g_{0} \, D_{00}^{2} \right\} n +O(\Omega^{0}).
\end{eqnarray}
To evaluate the excitation energies in a state with $f$ phonons I need the expectation value
\begin{eqnarray}
\< n ,f| : {\cal E}_{0}(\hat{n}_{0}) \, \hat{n}_{0} : |n, f\> = \overline{{\cal E}}_0 (n_0 -f )   ( n_0 - f)
\nonumber\\
\sim  \overline{{\cal E}}_0 (n_0) -  { \partial \over \partial n_0}
\left[  \overline{{\cal E}}_0 (n_0) \, n_0  \right]  f.
\end{eqnarray}
According to Eq.(\ref{fermionnumber2})  the coefficient of $f$ is of order $\Omega^{-1}$, so that to order $\Omega^0$ this term does not depend on f.

The  excitation energies of states containing $f $ phonons are therefore
\be
E_{K_{1}K_{2}...K_{f}}  =
  E_{K_{1}} +E_{K_{2}}...+E_{K_{f}}.
 \ee

 \section{Finite systems: The pairing model for nucleons in a single $j$-shell}
  
   I consider a system of nucleons  in a single $j$-shell, in which case the only fermion quantum number is  the third component of angular momentum $m$
  and the index of nilpotency  $\Omega $ is equal to $  j + { 1\over 2}$. Composite bosons are labelled by the angular momentum $L$ and its third component $M$:
 $K\equiv (L,M)$. The form factors of the potential are proportional to Clebsh-Gordan coefficients. 
 With the  normalization of Eq.\reff{potnormal} they are
  \be
  (F_{LM})_{m_1,m_2} = \sqrt{2  \Omega} \<jm_{1}jm_{2}|LM\>.
  \ee
   It is not necessary to solve the variational equations for the boson form factors, since because of rotational invariance they also must 
    be proportional to the  Clebsh-Gordan coefficients
  \be
   B_{LM}=\alpha_{L}F_{LM}.
   \ee  
   As a further simplification I restrict myself to the pairing model, namely I assume the single-particle energy to vanish and the interaction to be a pure 
   pairing potential  ($g_{K} = 0$ for $K \ne 0$ ). 
   I will evaluate the excitation energies to order $\Omega^0$ but the ground state energy only to order $\Omega$ since it was already evaluated
   to order $\Omega^0$ in ~\cite{Palu}.
 
  I can immediately find the normalizations
  \be
  \alpha_{0}= { 1 \over \sqrt { 1 -\nu}},\,\,\, \alpha_{L}= { 1\over 1 - \nu}, \,\,\, L \ne 0
  \ee
 and evaluate the coefficient of $H'$. I find that $\tilde{L} = L, \tilde{M}= -M$, and 
  \begin{eqnarray}
  E_C  &=& 2 h \Omega \nu^{2}  + \Omega g_0 \,\nu
  \nonumber\\
  {\cal E}_{0} &=& \left[ 2h - g_{0}(\Omega +2 +2 \nu\right]    ( 1-\nu) 
  \nonumber\\
 {\cal E}_{LM} &=&  \left[ 2h ( 1 - 2 \rho_{0})
  +  4 \Omega g_{0} \rho_{0} ( 1 -\rho_{0}) \right] 
\nonumber\\
{\cal V}_{LM} &=& (-1)^{M}  2 \left[  -2 h + \Omega g_{0} ( 1 - 2 \rho_{0})    \right]
\end{eqnarray}
where $\rho_0 = n_0/\Omega$.
Eqs.\reff{subtractionparameter} and \reff{fermionchemical} give
  \begin{eqnarray}
  \overline{r}^2 &=& n + O(\Omega^0)
  \nonumber\\
   \overline{\mu}_F &=& - { 1 \over 2}   g_{0} \left(  \Omega - n \right) +O(\Omega^{-1}).
   \end{eqnarray}
 With these values 
 \be
  {\cal V}_{LM}=O(\Omega^{-1}), \,\,\,{\cal E}_{LM}= \Omega g_{0} + O(\Omega^{-1}).
  \ee
 Since the number of L-modes is of order $\Omega^2$, namely $\delta E_C = O( \Omega^0)$,  the effective boson Hamiltonian  in the sector of $2n$ fermions is
 \begin{eqnarray}
 H_{B}(n) &=& \left[  - \Omega g_{0} n + g_{0}n^{2}  + O(\Omega^0) \right]
\nonumber\\
  & & + \left[\Omega g_{0}\sum_{LM} {\hat b}^{\dagger}_{LM} {\hat b}_{LM} + O(\Omega^{-1})  \right].
  \end{eqnarray}
This is the correct result~\cite{Ring} in  the required approximation.  It is to be noted that noncondensed bosons have all the same energy $E_{LM}= \Omega g_{0}$,
so that the spectrum depends only on their total  number, and not on their distribution in  different $L$-states. {\it This property makes
consistent the truncation of the boson space to a few modes}, and as a consequence  we can restrict the boson 
model space to only one of them, the  d-boson in the Interacting Boson Model.

The decoupling of different noncondensed bosons  is achieved by means of the subtraction of Eq.\reff{subtraction}.
With a subtraction independent of the structure functions all the modes are coupled, so that the above properties might have been found only after solution 
of the resulting boson Hamiltonian.

   \section{Infinite systems: Superconductors and the BCS model}
  
  I consider an infinite  system of  fermions whose quantum numbers are spin and momentum,  $m \equiv (s, p)$, and I parametrize the potential 
  form factors according to
  \be
 \left(F_{K} \right)_{s_{1}p_{1}s_{2}p_{2}} = \epsilon_{s_{1},s_{2}}  \delta_{p_{1} +p_{2},K} \, f_{p_{1}- { 1 \over 2}K}(K).
 \ee
    For simplicity I restrict myself to bosons of  spin zero and momentum $K$ and
    parametrize their form factors according to 
    \be
   (B_{K})_{s_{1}p_{1},s_{2}p_{2}} = \epsilon_{s_{1},s_{2}} \delta_{p_{1}+p_{2},K} \, \phi_{p_{1}- { 1 \over 2}K}(K).
   \ee
   Since form factor and structure function are fully antisymmetric 
  \be
  f_{q}(K) =  f_{-q} (K),  \,\,\, \phi_{q}(K)  = \phi_{- q}(K).
   \ee
   The structure functions are subject to the normalizations  of Eqs.\reff{ffnormal}
 \begin{eqnarray}
{ 1\over \Omega} \, \sum_{p}\Gamma_{p}  \, \phi_{p}^{2}  &=& 1
 \nonumber\\
 { 1\over \Omega}  \sum_{p}\, \Gamma^{2}_{ p} \, \phi^{2}_{p- { 1\over 2}K}(K)  &= &1, \,\,\, K \ne 0   \label{normBCS}
 \end{eqnarray}
 where
 \be
\phi_{p} = \phi_{p}(0) , \,\,\,\Gamma_{ p} = \left( 1 + \nu \, \phi_{p}^{2} \right)^{-1}.
 \ee
  For infinite systems  neglecting terms of order $\Omega^{0}$ in the evaluation of the ground state energy density does not produce any error
   in thermodynamic limit.  So I will neglect such terms. Therefore the ground state energy is  
 \begin{eqnarray}
E_{0} & = & \delta E_{C} + \left\{ {1\over \Omega} \mbox{tr} \left[ \Gamma B_{0} \, e \, B_{0}^{\dagger} \right] - 
\Omega g_{0} D_{00}^{2} \right\} n 
\nonumber\\
& = &\delta E_{C} + \left\{ \sum_{p} \, 2 \, e_{p}\,  \Gamma_{ p} \phi_{p}^{2} 
 - g_{0}\left[    \sum_{p}\Gamma_{ p} \phi_{p}f_{p} \right]^{2} \right\}  \,{ n \over \Omega}.
 \nonumber\\
  \end{eqnarray}
The expressions of  $ \delta E_{C} , {\cal E}_{K} $ and ${\cal V}_{K}$ can be easily evaluated inserting the definitions
of potential form factors and boson structure functions in the equations reported in  Appendix B. 

\subsection{BCS model}

For a general potential $ \delta E_{C} \ne O(\Omega^0) $, and  as already said  consistency of the  $\Omega^{-1}$ expansion requires  further subtractions. 
But a great simplification must arise for a pure pairing  interaction, $g_{K} =0$ for $K \ne 0$, because then as reminded previously  
the contribution of $\delta E_{C} $ to the energy per particle  must vanish in the thermodynamic limit, or in other words 
$\delta E_{C} $ must be of order $\Omega^{0}$.  Therefore I  assume $\delta E_{C}= O(\Omega^{0}) $ and I will verify a posteriori that this is true. 
 I also assume $\phi_{p}$ to be real and I minimize the ground state energy  imposing the normalization constraint by the  Lagrange multiplier $\lambda $
\be
{ \partial \over \phi_{q}}
\left\{ \sum_{p} 2 (e_{p}-\lambda) \Gamma_{ p} \phi_{p}^{2} 
 -  g_{0}\left[   \sum_{p}\Gamma_{p} \phi_{p}f_{p} \right]^{2} \right\}  =0.
\ee
I thus get the gap equation
\be
\left( e_{p} - \lambda \right)  \sqrt{\nu}\phi_{p} = { 1\over 2} \bigtriangleup  (1- \nu \, \phi^{2}_{p } )  f_{p}
\ee
where
\be 
\bigtriangleup = g_{0} \sqrt{\nu}\sum_{p} \Gamma_{p} \, \phi_{p} \, f_{p}
\ee
 is the gap function.  {\it These are exactly the results of the quasi-chemical equilibrium theory, Eq. IV(1.17)  of ref.~\cite{Blat}}. 
 In this connection I
 remind  that  this theory, unlike the BCS theory,  is fermion number conserving like the present approach.
 
 Following \cite{Blat} I then find
  $\lambda = e_{p_{F}}, p_{F}$ being the Fermi momentum of the system.
Next I  evaluate the coefficient of the quartic boson term
\begin{eqnarray}
{\cal V}_{K} &=& - { 2 \over \Omega}\sum_{p} \left\{ \Gamma_{ p}^{2} \, \Gamma_{ p - K}
 \left[2 (e_{p}- \overline{\mu}_{F}) \phi_{p} \right. \right.
 \nonumber\\
 &&
 \left. \left.  - { 1 \over \sqrt{\nu}}\left(1 -  \nu \phi_{p}^{2}  \right)
 \bigtriangleup f_{p} \right]  \phi_{p - K}\right.
\nonumber\\
&&\left.
  \times    \phi_{p- { 1\over 2}K}(K) \phi_{p- { 1\over 2}K}(-K)\right\} + O(\Omega^{-1}).
\end{eqnarray}
In the above equation there appears the fermion chemical potential $\overline{\mu}_{F}$ which is determined by Eq.(\ref{fermionchemical}). 
Using the gap equation  I find $  \overline{\mu}_{F} = \lambda $. 
Then using  again the gap equation  in the expression of  $ {\cal V}_{K}$ I find  that  $ {\cal V}_{K}= O(\Omega^{-1}) $ and therefore 
$\delta E_{C }$ is at most  $O(\Omega^0) $. 
{\it For a pure pairing interaction the contribution to the ground state energy per particle of noncondensed pairs vanishes,
 in agreement with ~\cite{Haag}, and my assumption about $\delta E_{C }$ is justified}. 

To get  analytic expressions  I assume  $e_p$ to be the free fermion energy and the form factor of the potential to have 
the schematic form of BCS
 \be
  e_{p}= { 1 \over 2m}p^{2}, \,\,\, f_{p}= \theta( 2m \omega - | p^{2} - p_{F}^{2}| ).
   \ee
 $\theta$ is the step function and $\omega$ a cutoff energy (identified with the Debye cutoff energy of lattice waves
 in metal superconductivity ).
 
The solution of the gap equation is~\cite{Blat}
\be
\phi_{p}={ 1 \over \nu} \left( \sqrt{1 + \xi_{p}^{2}} - \xi_{p}\right)
\ee
where
\be
\xi_{p}= { 1\over \bigtriangleup}{ 1 \over 2\, m}(p^{2} - p_{F}^{2}).
\ee

\section{Summary and outlook}

I extended the original formalism for  boson dominance  based on coherent composite states.
The extension amounts to perform a subtraction proportional to  the structure function  of the condensed boson. 
The results obtained have  a  validity restricted only by the assumption that condensation occurs in a singlet state and that
the coupling constants scale according to $g_{K} \sim  \Omega^{-1}$. In particular, since fermion number is conserved, they 
hold for finite systems. Therefore the present bosonization method can find applications  not only in the presence of
important nonpairing interactions, but also to account for finiteness effects in atomic nuclei and small metallic grains.

 The general expression of the bosonic Hamiltonian contains a classical part  $ \delta  E_C$ which couples the form factors of all the bosons. 
 The theory becomes much simpler for pure pairing interactions, because $ \delta  E_C$ is negligible,
 and I tested it on two paradigmatic models. 
 
 As an example of
finite systems I considered the pairing model  of nucleons in a single $j$-shell and I reproduced  its bosonic
spectrum in a form useful to understand and justify the Interacting Boson Model. 
{\it Indeed I find that a boson space made of the $s$- and $d$-bosons is sufficient to reproduce the spectrum, which
is not changed by the addition of other bosons}. For the fermionic part of the spectrum one must use
the technique developed in~\cite{Paluf}.
Needless to say the bosonization of the pairing model has been obtained in many ways, but I emphasize that the present formulation holds
in a framework valid for arbitrary fermion-fermion interactions.
 
I then applied this method to the  BCS model of superconductivity reproducing exactly its  ground state properties in the form of the quasi-chemical equilibrium theory.

Two important issues are left for future work. The first one concerns the determination of structure functions for a general interaction and 
noncondensed bosons. I keep in mind in this connection the possibility of different structure functions with the same quantum 
number $K$. They might for instance be  associated with intruders or coexisting molecular and Cooper pairs.

The second one concerns phonons, namely neutral bosons describing polarization effects~\cite{Gor}. They
 must necessarily be included  in many cases~\cite{Heis}  and certainly in the presence of 
particle-hole terms in the fermion-fermion interaction.

\appendix

\section{Basic formulae in Berezin integrals \label{Berezin}}

The definition of the Berezin integral for a single Grassmann variable is
\be
\int d \gamma ( a \gamma + b) = a,
\ee
the generalization to many variables being obvious. For a change of variables
\be
\gamma_i = \gamma_i ( \gamma')
\ee
in a multiple integral we have
\be
\int  \prod_i (d \gamma_i) f ( \gamma) = \left( \det {\partial \gamma_h \over \partial \gamma_k'} \right)^{-1}
\int \prod_i (d \gamma_i') f( \gamma').
\ee
Notice the appearance of the inverse of the jacobian, contrary to the case of ordinary variables.

Gaussian integrals can be evaluated exactly, like for ordinary variables. There are two types of such integrals
\begin{eqnarray}
\int \prod_h (d \gamma_h^* d \gamma_h) \exp \sum_{ij} \gamma_i^* M_{ij} \gamma_j &=& \det M
\\
\int \prod_h (d \gamma_h) \exp \sum_{ij} { 1 \over 2} \gamma_i A_{ij} \gamma_j &=& \mbox{Pf} \,A  \label{pfaffian}
\end{eqnarray}
where $\mbox{Pf} \,A$ is called the $pfaffian$~\cite{Zinn} of $A$. The following algebraic identity holds
\be
\left( \mbox{Pf} \,A \right)^2 = \det A.
\ee

\section{Inner products of composite states \label{inner} }

Let us consider the case of only one composite. To evaluate the inner product of coherent states
I use the identity operator in the fermion Fock space
\be
{\cal I} = \int d \gamma^* d \gamma \<\gamma | \gamma\>^{-1} | \gamma \> \<  \gamma |
\ee
where the $\gamma^*,\gamma$ are Grassmann variables and $|\gamma \>$ coherent  states~\cite{Zinn}
\be
|\gamma \> = \exp (- \gamma \, {\hat c}^{\dagger}) \>.
\ee
I then have
\be
\<\beta_1|\beta\> = \<\beta_1| {\cal I} |\beta\>= \int d \gamma^* d \gamma \, \exp ( - \gamma^*  \gamma ) 
\<\beta_1 | \gamma\> \< \gamma |\beta\>.
\ee 
The matrix element $ \<\beta_1| \gamma\>$ can be evaluated  using the defining property of coherent states
\be
{\hat c} | \gamma\> = \gamma| \gamma\>
\ee
with the result
\be
\<\beta_1|\gamma\> = \exp \left( { 1\over 2 \sqrt{\Omega}} \beta_1^* \gamma B \gamma \right).
\ee
Therefore $\<\beta_1|\beta\> $ becomes 
\be
\<\beta_1|\beta\> = \int d \gamma^* d \gamma \, E(\gamma^*,\gamma,\beta_1^*,\beta),
\ee
where the function $E$ is 
\begin{eqnarray}
 E(\gamma^*,\gamma,\beta^*,\beta)   & = & \exp \left( -\gamma^* \gamma + 
{1\over 2 \sqrt{\Omega}} \beta^* \, \gamma \, B \, \gamma \right.
\nonumber\\
 & & \left. +  { 1\over 2\sqrt{\Omega}} \beta  \, \gamma^* B^{\dagger} \gamma^* \right). \label{E}
\end{eqnarray}
 By the change of variables 
\be
\gamma' =  \gamma^* - { \sqrt{\Omega} \over \beta} (B^{\dagger})^{-1} \gamma
\ee
the integral is factorized according to
\begin{eqnarray}
& & \<\beta_1| \beta\> = \int \, d \gamma' \exp \left({ 1 \over 2 \sqrt{\Omega}} \gamma' \,  \beta\,
B^{\dagger}
\,
\gamma'
\right) \times 
\nonumber\\ 
 & &  \int  d \gamma  \exp \left[ { 1\over 2} \gamma \left( \sqrt{\Omega} \left( \beta 
B^{\dagger}\right)^{-1} + { 1 \over \sqrt{\Omega}}\beta_1^*  B \right)  \gamma \right].
\end{eqnarray}
The factors are of the form~\ref{pfaffian}, so that finally
\begin{eqnarray}
& &\<\beta_1|\beta\> = \left[ \det \left( { 1 \over \sqrt{\Omega}}\beta B^{\dagger}\right) \right]^{{ 1 \over 2}}  
\left[ \det \left(   \sqrt{\Omega}  ( \beta B^{\dagger} )^{-1} + \right. \right.
\nonumber\\
& & \,\,\,\,
  \left. \left.{ 1 \over \sqrt{\Omega}}\beta_1^* B \right) \right]^{{1 \over 2}} = \det \left[ 1\!\!1 + { 1 \over
\sqrt{\Omega}}\beta \, \beta_1^* B^{\dagger} B \right] ^{{1 \over 2}}.
\end{eqnarray}

\section{ The operator $\mathcal{P}$ \label{projector}}

In this Appendix I show that the operator $ \mathcal{P} $ of Eq.\ref{Poperator} approximates  
$ \mathcal{P}_C  $ to leading order in an expansion in the inverse of the index of nilpotency.

For the sake of simplicity I consider the case of a unique composite. 
Then
\begin{equation}
 \mathcal{P}_C  = \sum_{n =0}^{\Omega}  {1 \over  \nu_n } \, 
  | \left(\hat{B}^\dagger \right)^n |0 \rangle \langle 0| \hat{B}^n |
\end{equation}
where
\begin{equation}
 \nu_n  =  \langle 0| \hat{B}^n | \left( \hat{B}^\dagger \right)^n|0 \rangle \,.
\end{equation}
I must then prove that
\begin{equation}
\langle 0 | \hat{B}^m \, \mathcal{P} ( \hat{B}^\dagger )^n |0\rangle 
\sim  \langle 0 | \hat{B}^m |  ( \hat{B}^\dagger)^n |0\rangle = \delta_{m,n} \nu_m \,.\label{Papp1}
\end{equation}
These equations are generated by the following ones
\begin{equation}
\langle \beta' | \mathcal{P} |\beta" \rangle \sim \langle \beta' | \beta" \rangle \label{gen}
\end{equation}
taking derivatives with respect to $\beta'^*, \beta" $ and setting these variables equal to zero.
To simplify the formulae I adopt a slightly different definition of composites
\begin{equation}
\hat{B} = { 1\over 2} \sum _{m_1 m_2} {\hat c}^{\dagger}_{m_1} B_{m_1 m_2}{\hat c}_{m_2}.
\end{equation}
The right and left hand sides of  Eq. \ref{gen} are
\begin{eqnarray}
\langle \beta' | \mathcal{P} | \beta'' \rangle & =& \int { d \beta^* d\beta \over 2 \pi i} 
\exp \left[ -\mathcal{E}(\beta^*,\beta, \beta'^*, \beta'') \right]
\nonumber\\
\langle \beta' | \beta" \rangle &=& \exp \left[ \mbox{Tr} \ln (1 + \beta'^* \beta" B B^\dagger) \right]
\end{eqnarray}
where
\begin{eqnarray}
& & \mathcal{E}(\beta^*,\beta, \beta'^*, \beta'') = \mbox{Tr} \left[  \ln (1 + \beta^* \beta  B B^{\dagger}) 
\right.
\nonumber\\
&  & \left.  -  \ln (1  + \beta'^* \beta  B B^{\dagger}) - \ln (1 + \beta^* \beta''  B B^{\dagger})  \right] \,.
\end{eqnarray}
I evaluate the integral by the saddle point method.  The saddle point equations are
\begin{eqnarray}
(\beta - \beta'') \mbox{Tr} { B B^{\dagger} \over (1 + \beta^* \beta B B^{\dagger} )
(1 + \beta^* \beta'' B B^{\dagger} )} &=&0
\nonumber\\
(\beta^* - \beta'^*) \mbox{Tr} { B B^{\dagger} \over (1 + \beta^* \beta B B^{\dagger} )
(1 + \beta'^* \beta B B^{\dagger} )} &=&0  
\end{eqnarray}
with solutions
\begin{equation}
\overline{\beta} = \beta'',  \,\,\, \overline{\beta}^* = \beta'^* \,.
\end{equation}
At the saddle point
\begin{equation}
\mathcal{E}(\overline{\beta}^*,\overline{\beta}, \beta'^*, \beta'') =  
-  \mbox{Tr} \ln (1 + \beta'^* \beta''  B B^{\dagger})  \,.
\end{equation}
Moreover
\begin{eqnarray}
& &\frac{\partial^2 {\cal E}}{\partial \beta^* \partial \beta^*}|_
{\overline{\beta} = \beta'',\overline{\beta}^* = \beta'^* } =
\frac{\partial^2 {\cal E}}{\partial \xi \partial \xi}|_{\overline{\beta} = 
\beta'',\overline{\beta}^* = \beta'^* } =0
\nonumber\\
& &\frac{\partial^2 {\cal E}}{\partial \beta^* \partial \beta}|_
{\overline{\beta} = \beta'',\overline{\beta}^* = \beta'^* }= \mbox{Tr} {  B B^{\dagger}
\over (1 + \beta'^* \beta''  B B^{\dagger})^2} \,.
\end{eqnarray}
In conclusion
\begin{equation}
\langle \beta' | \mathcal{P} | \beta" \rangle \sim \langle \beta' | \beta" \rangle  \, \left[ \mbox{Tr} { B B^{\dagger}
\over (1 + \beta'^* \beta''  B B^{\dagger})^2} \right]^{-1}\,.
\end{equation}
The desired result follows if we assume
\begin{equation}
\mbox{Tr} \left( B^\dagger B \right)^n \sim \Omega^{-n+1} \,.
\end{equation}
It is then easy to prove also  the idempotency property of projectors 
\begin{equation}
 \mathcal{P} \sim    \mathcal{P}^2 \,.
\end{equation}

\section{Derivation of the effective action \label{action}}

 For the following manipulations we need the Hamiltonian in antinormal form
\be
H = { 1\over 2} \mbox{tr} ( h + h_0 ) - {\hat  c} \,  h^T  {\hat c}^{\dagger} -
\sum_K g\K \,  {1\over 2} \, {\hat c} F_K {\hat c}
\, { 1\over 2} \, {\hat c}^{\dagger} F^{\dagger}_K {\hat c}^{\dagger}
\ee
where the upper script $T$  means "transposed" and $h$ was given in Eq.(\ref{hoperator}).
Now we must evaluate the matrix element $\< \beta_t| \exp(- \tau  H) |\beta_{t-1} \>$. To this end we expand to 
first order in
$\tau$ (which does not give any error in the final $\tau \rightarrow 0$ limit) and
insert the operator ${\cal P}$ between annihilation and creation operators
\begin{eqnarray}
& &\< \beta_t| \exp ( - \tau H ) |\beta_{t-1}\> = \exp \left( - {1\over 2} \mbox{tr}(h+h_0) \tau \right) 
\,\,\<\beta_t| {\cal{P}}
\nonumber\\
& & \,\,\,\,\, 
 - {\hat  c} \,  h^T \tau \, {\cal{P}} {\hat c}^{\dagger}
 \sum_k g_k \tau  \, {1\over 2} \, {\hat c} F_K{\hat  c} \, {\cal{P}}
\, { 1\over 2} \,{\hat  c}^{\dagger} F^{\dagger}_K{\hat  c}^{\dagger} |\beta_{t-1}\>.
\end{eqnarray} 
Using the identity in the fermion Fock space we find
\begin{eqnarray}
& &\<\beta_t|  \exp (- \tau H ) |\beta_{t-1}\> =  \int d \gamma^* d \gamma  \, 
E(\gamma^*,\gamma,\beta_t^*,\beta_{t-1})
\nonumber\\
& & \,\,\,\,\,\,\,\,\,\, \times   \exp\left( - {1\over 2} \mbox{tr}(h+h_0) \tau   - \gamma^* h \, \tau \gamma \right)
\nonumber\\ 
& & \,\,\,\,\,\, \,\,\,\,\times 
\exp \left(   \sum_K g_K \tau \, {1\over 2} \gamma \,F_K \,\gamma  
 \, { 1\over 2}  \gamma^* F_K^{\dagger} \gamma^* \right)
\end{eqnarray}
where the function $E(\gamma^*,\gamma,\beta^*,\beta)$ is defined in~(\ref{E}).
By means of the Hubbard-Stratonovich transformation we can make the exponents quadratic in the Grassmann
variables and evaluate the Berezin integral
\begin{eqnarray}
& & \<\beta_t| \exp(- \tau  H) |\beta_{t-1} \> =  \int \prod_K da_K^* da_K \exp (   - a^* \cdot a  )
\nonumber\\
& & \,\,\, \times \exp \left\{ 
 { 1\over 2} \mbox{tr} \ln \left[ 1\!\!1 + \left( {\mathcal B}_t^* + \sum_{K_1} \sqrt{ g_{K_1}
\tau} 
\,
 a^*_{K_1} F_{K_1}\right)
\right. \right.
\nonumber\\
& & \,\,\, \left. \left.  \times R^{-1} \left( {\mathcal B}_{t-1} + \sum_{K_2} \sqrt{ g_{K_2} \tau} 
\, a_{K_2} (F_{K_2})^{\dagger}\right) (R^T)^{-1} \right] \right\}
\nonumber\\
& & \,\,\, \times  \det \rho \, \exp \left( - {1\over 2} \mbox{tr}(h+h_0) \tau \right),
\end{eqnarray}
where
\be
\rho = 1\!\!1 + h \, \tau.
\ee
Performing the integral over the auxiliary fields $a\K^*,a\K$ we get 
\begin{eqnarray}
& & \<\beta_t| \exp(- \tau  H) |\beta_{t-1} \> =  \int \prod_K da_K^* da_K \exp (   - a^* \cdot a  )
\nonumber\\
& & \,\,\, \times \exp \left\{ 
 { 1\over 2} \mbox{tr} \ln \left[ 1\!\!1 + \left( {\mathcal B}_t^* + \sum_{K_1} \sqrt{ g_{K_1}
\tau} 
\,
 a^*_{K_1} F_{K_1}\right)
\right. \right.
\nonumber\\
& & \,\,\, \left. \left.  \times  \rho^{-1} \left( {\mathcal B}_{t-1} + \sum_{K_2} \sqrt{ g_{K_2} \tau} 
\, a_{K_2} (F_{K_2})^{\dagger}\right) (\rho^T)^{-1} \right] \right\}
\nonumber\\
& & \,\,\, \times  \det \rho \, \exp \left( - {1\over 2} \mbox{tr}(h+h_0) \tau \right),
\end{eqnarray}
The functional form of the composites partition function is
\be
Z_C = 
 \int \left[{d \beta^* d \beta \over 2 \pi i} \right]\, \exp \left(-  S_{\mbox{eff}}(\beta^*,\beta) \right)
\ee
where $ S_{\mbox{eff}}$ is given in Eq.~(\ref{bosaction}).

\section{General form of the coefficients in the auxiliary Hamiltonian \label{coefficients}}
 
  I will use the definitions
  \begin{eqnarray}
  T_{K_{1}K_{2}...K_{2l-1}K_{2l}}& =& { 1\over  \Omega} \mbox{tr}\left[  C_{K_{1}K_{2}}... 
  C_{K_{2l-3}K_{2l-2} }\right. 
  \nonumber\\
  && \left.
    \times \Gamma B_{K_{2l-1}} \, h \, B^{\dagger}_{K_{2l}}  \right]
  \nonumber\\
   D_{K_{1}K_{2}...K_{2l-1}K_{2l}}& =& { 1\over 2 \Omega} \mbox{tr}\left[  C_{K_{1}K_{2}} ...   C_{K_{2l-3}K_{2l-2} }\right. 
   \nonumber\\
   & & \left.   
  \times  \Gamma B_{K_{2l-1}} F_{K_{2l}}^{\dagger} \right].
   \nonumber\\
    \end{eqnarray}
    for $l$ an arbitrary integer. Let me emphasize that only the last factor in $T$ involves the single particle kinetic energy $h$, and only the
    last factor in $D$ involves the form factors of the potential. {\it In the following I neglect the coupling to external fields and
    I assume that if only 2 indices $K_1,K_2$ are different from zero, they must be equal in the $ {\mathcal E}$- terms}.
  
 The classical energy is
 \begin{eqnarray}
   E_C  &=&  \Omega  \, \nu^{2}T_{0000} + \sum_{K}  \Omega g_{K} \nu
   { 1 \over 2 \Omega} \mbox{tr} \left[C_{00} F_{K}^{\dagger}F_{K} \right]
   \end{eqnarray}
 The $s$-boson energy  is given by
  \begin{eqnarray}
 {\cal E}_{0} &=&  T_{00} -  \nu  \, T_{0000} + \Omega g_{0}\left[ - |D_{00}|^{2} + 2 \left({ \hat{n}_{0}\over \Omega} - \nu \right)  \right.
  \nonumber\\
    && \left.  \times  D_{00}D_{0000}\right] + 
  \nonumber\\
    & &
  + \sum_{K }   g_{K} 
   { 1 \over 2 \Omega} \mbox{tr} \left[\Gamma B_{0} F_{K}^{\dagger} \Gamma F_{K}B_{0}^{\dagger} \right] .
  \end{eqnarray}
  The terms ${\cal E}, {\cal V}$ are separated in their contributions from the kinetic and potential terms in the fermion Hamiltonian
  \begin{eqnarray}
  {\cal E}_{K_{1}K_{2}} &=&   {\cal E}_{K_{1}K_{2}}^{\mbox{kin}} + {\cal E}_{K_{1}K_{2}}^{\mbox{pot}} 
\nonumber\\
   {\cal V}_{K_{1},K_{2}}&=& {\cal V}_{ K_{1},K_{2}}^{\mbox{kin}}+ {\cal V}_{ K_{1},K_{2}}^{\mbox{pot}} 
  \end{eqnarray}
  whose expressions are
    \begin{eqnarray}
     {\cal E}_{ K_{1},K_{2}}^{\mbox{kin}}  &=&  T_{K_{1}K_{2}} - \nu  \, T_{ K_{1}K_{2}00}  
    -   \left[T_{K_{1}00K_{2}} +T_{0K_{2}K_{1}0}\right.
    \nonumber\\
   & &  \left.  - \nu  ( T_{0K_{2}K_{1}000} + T_{K_{1} 0 0 K_{2}00} \right] { n \over \Omega}
   \nonumber\\
   {\cal E}_{K_{1}K_{2}}^{\mbox{pot}} & = & 2 \left\{ g_{0} D_{00} \left[  D_{K_{2}K_{1}0 0} +D_{0 K_{1}K_{2}0} 
  - { \hat{n}_{0} \over \Omega} \left(  D_{0 K_{1}K_{2}0 0 0} \right. \right.  \right.
\nonumber\\
& & \left. \left.  \left. + D_{K_{2}0 0 K_{1}0 0} \right) \right]^{\dagger}  \hat{n}_{0}+
\sum_{K \ne 0} \hat{n}_{0} \,  g_{K} \left[  D_{K_1 K}D_{K_2 0 0 K}^{\dagger} \right. \right.
 \nonumber\\
 & &
\left.   \left. - { \hat{n}_0 \over 2 \Omega}  \left( D_{0 K_1 0K}^{\dagger} D_{0 K_2 0  K}
  + D_{ K_1 0 0K} D_{ K_2  0 0  K}^{\dagger} \right) \right]  \right\}
  \nonumber\\
  \nonumber\\
     {\cal V}_{K_{1}K_{2}}^{\mbox{kin}} & =& - 2 \left[ T_{K_{1}0K_{2}0} - \nu  \, T_{K_{1}0K_{2}000} \right] 
  \nonumber\\
  \nonumber\\
   {\cal V}_{K_{1}K_{2}}^{\mbox{pot}} & =&  2 \left\{ \Omega g_{0} D_{00} \left[ D_{K_{2} 0 K_{1}0} - { \hat{n}_{0} 
   \over \Omega} \left( D_{0K_{1}0 K_{2}0 0} \right. \right. \right.
  \nonumber\\
  &  & \left.  \left.  \left.+
   D_{K_{1}0 K_{2}0 0 0} \right)^{\dagger} \right]
   + \sum_{K\ne 0} \Omega g_{K} \left[ D_{K_{1}K} D_{0K_{2}0K}^{\dagger} \right. \right.
   \nonumber\\
   & & \left. \left.
    - { \hat{n}_{0} \over \Omega} D_{0K_{1}0K}^{\dagger}D_{K_{2}00K}   \right] \right\}.
      \end{eqnarray}

%\clearpage 

\end{document}